\documentclass[preprint,prl,showpacs]{revtex4}
\usepackage{graphicx}
\usepackage{dcolumn}
\usepackage{bm}

\begin{document}

\author{R. K\"uchler$^{(1)}$, N. Oeschler$^{(1)}$, P. Gegenwart$^{(1)}$,
T. Cichorek$^{(1)}$, K. Neumaier$^{(2)}$, O. Tegus$^{(3)}$, C.
Geibel$^{(1)}$, J.A. Mydosh$^{(1),(4)}$, F. Steglich$^{(1)}$, L.
Zhu$^{(5)}$, and Q. Si$^{(5)}$}

\address{$^{(1)}$Max-Planck Institute for Chemical Physics
of Solids, D-01187 Dresden, Germany
\\ $^{(2)}$Walther Meissner Institute, D-85748 Garching, Germany
\\ $^{(3)}$ Van der Waals-Zeeman Laboratory, University of
Amsterdam, The Netherlands
\\ $^{(4)}$Kamerlingh Onnes Laboratory, Leiden University, The Netherlands
\\ $^{(5)}$Department of Physics and Astronomy, Rice University,
Houston, TX 77005-1892}

\title{Divergence of the Gr\"uneisen Ratio at Quantum Critical
Points in Heavy Fermion Metals}

\begin{abstract}

We present low-temperature volume thermal expansion,
$\beta$, and specific heat, $C$,
measurements on high-quality single crystals of CeNi$_2$Ge$_2$ and
YbRh$_2$(Si$_{0.95}$Ge$_{0.05}$)$_2$ which are
located very near to quantum critical points. For both systems,
$\beta$ shows a more singular temperature
dependence than $C$, and thus the Gr\"uneisen ratio
${\Gamma\propto\beta/C}$ diverges as $T\rightarrow 0$. For
CeNi$_2$Ge$_2$, our results are in accordance with the
spin-density wave (SDW) scenario for three-dimensional
critical spin-fluctuations. By contrast, the observed singularity
in YbRh$_2$(Si$_{0.95}$Ge$_{0.05}$)$_2$ cannot
be explained by the itinerant SDW theory but is qualitatively
consistent with a locally quantum critical picture.

\end{abstract}

\pacs{71.10.HF,71.27.+a}
\maketitle

Quantum critical points (QCPs) are of extensive current interest
to the physics of correlated electrons, as proximity to a QCP
provides a route towards non-Fermi liquid behavior. While a broad
range of correlated electron materials are being studied in this
context,
heavy fermions have been playing an especially important
role:
a growing list of heavy fermion (HF) metals explicitly
displays magnetic QCPs \cite{Stewart,Schroeder,Lonzarich,Gegenwart CNG,%
Trovarelli Letter}. Systematic experiments in these systems
promise to shed considerable light on the general physics of
quantum critical metals.
Indeed, recent experiments
\cite{Stewart,Schroeder} have shown that, at least in some of the
HF metals, the traditional theory of metallic magnetic quantum
phase transition fails. This traditional picture
\cite{Sachdev-book} describes a $T=0$
SDW
transition and, relatedly, a mean-field type of quantum critical
behavior. More recently, it has been shown that a destruction of
Kondo resonances can lead to a breakdown of the SDW picture
\cite{Sietal,Colemanetal}; what emerges instead are new classes
of QCPs that are locally critical \cite{Sietal}.

Given these experimental and theoretical developments, it seems
timely to address the conditions under which these different types
of QCPs arise. For this purpose, it would be important to carry
out comparative studies of different heavy fermion materials.
This paper reports one such study.
We have chosen the HF systems
CeNi$_2$Ge$_2$ \cite{Gegenwart CNG} and
YbRh$_2$(Si$_{0.95}$Ge$_{0.05}$)$_2$ \cite{Trovarelli
Letter,Custers Nature}, both of which
crystallize in
the tetragonal ThCr$_2$Si$_2$ structure. Both are
ideally suited to study antiferromagnetic (AF) QCPs since they
are located very near to the magnetic instability,
and since the effect of disorder is minimized in these high quality
single crystals with low residual
resistivities.
We have
focused on measurements of the thermal expansion, $\beta$,
and Gr\"uneisen ratio, $\Gamma\propto\beta/C$, where
$C$ denotes the specific heat, since recent theoretical work
\cite{Zhu} has shown that i) $\Gamma$ is divergent as $T$ goes to zero
at any QCP and ii) the associated critical exponent can be used
to differentiate between different types of QCP.

Presently measurements of the thermal
expansion and Gr\"uneisen ratio for systems located directly at the QCP
are lacking. Only for
Ce$_{1-x}$La$_x$Ru$_2$Si$_2$, which orders antiferromagnetically
for ${x>x_c}$ with ${x_c=0.075}$, $\beta$ has
been measured for concentrations ${x=0}$ and ${x=0.05}$ at temperatures
above 0.4 K. A very large $\Gamma$ was
obtained which, however, was found to saturate at low
temperatures \cite{Kambe}.
In other 
solids
too,
all previous measurements reported
in the literature yield a finite Gr\"uneisen
ratio \cite{Barron}.

In this Letter we communicate the first-ever observation of a divergent Gr\"uneisen ratio ${\Gamma}$ for
${T\rightarrow 0}$. CeNi$_2$Ge$_2$ is known to be a NFL compound which exhibits a paramagnetic ground state
\cite{SC CNG}. The electrical resistivity, $\rho(T)$, resembles that of CePd$_2$Si$_2$ at the pressure tuned QCP
\cite{Grosche}: $\rho-\rho_0\propto T^\epsilon$ with $1.2 \leq\epsilon\leq1.5$ below 4 K \cite{Gegenwart
CNG,Grosche,Gegenwart Physica,Koerner,Braithwaite}. In YbRh$_2$Si$_2$ pronounced NFL effects, i.e.
$C/T\propto-\log(T)$ and $\Delta\rho\sim T$ have been observed upon cooling from 10 K down to 0.3 K. While
$\Delta\rho(T)$ keeps following the linear $T$-dependence down to $T_N=$70 mK, $C/T$ diverges stronger than
logarithmically below $T=0.3$ K \cite{Trovarelli Letter,Gegenwart YRS} . For our study we chose a high-quality
single crystal of YbRh$_2$(Si$_{0.95}$Ge$_{0.05}$)$_2$ for which $T_N$ has been reduced to 20 mK. Large
CeNi$_2$Ge$_2$ single crystals of weight 5 to 6 g were grown using the traveling-floating-zone mirror-oven
technique. The samples were analyzed via electron-probe microanalysis (EPMA) and found to have the desired 122
(P4/mmm) structure with little mosaic spread, good stoichiometry and no second phases. Two thin oriented bars
with a length of 5 mm were formed by spark-erosion from the center of the large single crystal. The bars were
annealed for 120 hours at 800$^\circ$C in an Ar partial pressure ($10-20$ torr) and have a residual resistivity
of $3-5~\mu\Omega$cm. Single crystalline platelets of YbRh$_2$(Si$_{1-x}$Ge$_x$)$_2$ with a nominal Ge
concentration of $x=0.05$ were grown from In flux as described earlier \cite{Trovarelli Letter,Custers Nature}.
From a careful EPMA,
the
${\it effective}$ Ge-concentration
is found to be
$x_{eff}\leq0.02\pm0.01$. The large difference between nominal
and effective Ge-content is due to the fact, that Ge dissolves
better than Si in the In-flux. A similar effective
Ge content of ${0.02\pm0.004}$ \cite{Custers Nature} is deduced
from hydrostatic pressure experiments
\cite{Mederle}.
The residual resistivity of the Ge-doped crystal is 5 $\mu\Omega$cm.
The thermal expansion and the
specific heat have been determined in dilution refrigerators by
utilizing an ultrahigh resolution capacitive
dilatometer and the quasi-adiabatic heat pulse technique, respectively.

Figure~1 displays the $T$ dependence of $\alpha_a$ and $\alpha_c$, the linear thermal expansion coefficients of
CeNi$_2$Ge$_2$ measured along the tetragonal $a$- and $c$-axis. As shown by the solid lines, the data can be
described in the entire $T$ range investigated by the $T$-dependence predicted \cite{Zhu} by the
three-dimensional (3D) SDW scenario, i.e., the sum of (singular) square-root and (normal) linear contributions.
The corresponding fit parameters are listed in Table I. We observe a moderate anisotropy $\alpha_c\simeq
1.8\alpha_a$.
As shown in the inset, the volume expansion coefficient
$\beta=2\alpha_a+\alpha_c$, plotted as $\beta(T)/T$ is not a constant upon cooling, as
would be
for a Fermi liquid,
but shows a $1/\sqrt{T}$ divergence over more than two decades
in temperature from 6 K down to at least 50 mK.
This is one of the cleanest observations of NFL behavior in
a thermodynamic property made in any system so far.

We next consider the low temperature specific heat of CeNi$_2$Ge$_2$.
As shown by several investigations, $C(T)/T$
strongly increases upon cooling from 6 K to 0.4 K
\cite{Knopp,Aoki,Gegenwart CNG,Koerner,Steglich2000}. This
increase has either been described by $C(T)/T\propto-\log(T)$
\cite{Gegenwart CNG,Koerner} or
$C(T)/T=\gamma_0-c\sqrt{T}$ \cite{Aoki}. Below 0.4 K different
behaviors have been reported. While Knopp {\it et
al.} found a peak at 0.3 K followed by a 6\% decrease in $C(T)/T$
from the maximum value \cite{Knopp}, Koerner
{\it et al.} observed a leveling off in $C(T)/T$ below 0.3 K
\cite{Koerner}. In contrast, $C(T)/T$ of a
high-quality sample with very low residual resistivity does not
saturate but shows an upturn at the lowest
temperatures \cite{Steglich2000}. Very recently a systematic study
of the low temperature specific heat on
different high-quality polycrystals, prepared with a slight
variation of the stoichiometry \cite{Gegenwart
Physica}, has been performed. The result was that nearly all
of the different investigated samples showed such an
upturn in $C(T)/T$ below 0.3 K whose size, however, is strongly
sample dependent even for samples with similar
$\rho(T)$ and a residual resistivity of only 0.2 $\mu\Omega$cm
\cite{Cichorek}. In the following we analyze the
specific heat (Fig.~2) measured on the same sample that has been
used for the thermal expansion study. Below 3 K
the data can be described by $C(T)/T=\gamma_0-c\sqrt{T}+d/T^3$
using the parameters listed in Table I (solid
lines in Fig.~2). Here we assume that the low temperature upturn,
present in this single crystal as well, could
be ascribed to the high-temperature tail of a Schottky anomaly
\cite{Cichorek_tbp}. Its influence on the Gr\"uneisen ratio is
smaller than $5\%$ at 0.1 K and therefore not
visible in the $\Gamma(T)$ plot shown in the inset of Fig.~2.
This is the first observation of a divergent
$\Gamma(T)$ for $T\longrightarrow 0$ in any material and provides
striking evidence that CeNi$_2$Ge$_2$ is
located very close to a QCP. The observed $T$ dependence is in full
agreement with the 3D SDW prediction
\cite{Zhu}. If the investigated high-quality single crystal would
enter a Fermi liquid regime below 0.3 K as
observed for the sample studied in \cite{Koerner}, $\Gamma(T)$
should saturate below that temperature.

The application of magnetic fields to CeNi$_2$Ge$_2$ is found to
gradually reduce the low-$T$ specific heat
coefficient. For $B\geq 2$ T a nearly
temperature-independent
$\gamma(B)=C(T,B)/T$
is observed at low temperatures with
$\gamma(B)=\gamma_0-const\sqrt{B}$ \cite{Aoki}.
The low-temperature thermal expansion shows a similar field-induced
crossover to Fermi liquid behavior (Fig.~3)
and the field dependence of $\alpha(T,B)/T$ in the field induced FL
regime diverges like $1/\sqrt{B}$ (not shown).
Both features are consistent with
the predictions \cite{Zhu} from the
itinerant 3D SDW fluctuations at a zero-field AF QCP,
assuming a linear
dependence between the magnetic field and the distance
$r$ from the QCP.

We now turn to YbRh$_2$(Si$_{0.95}$Ge$_{0.05}$)$_2$, in which we have
measured the thermal expansion from 50 mK to
6 K. Compared to CeNi$_2$Ge$_2$, here the volume thermal expansion
coefficient $\beta(T)$ has an opposite sign
reflecting the opposite volume dependence of the characteristic energies.
At $T>1$ K, $\beta(T)$ can be fit by
$-T \log (T_0/T)$ with $T_0\approx 13$ K (see Fig.~4). At $T<1$ K,
the best fit is given by $a_1+a_0T$. Both are
not only different from the expected 3D-SDW results discussed earlier,
but also weaker than the $\ln \ln T$ form
\cite{Zhu} expected in a 2D-SDW picture \cite{YRS beta}. The difference
from the 2D-SDW picture is even more
striking when we look at the Gr\"uneisen ratio. In Fig.~4, we have also
shown the electronic specific heat at
zero magnetic field. Here $C_{el}=C-C_Q$, where $C_Q\propto1/T^2$ denotes
the nuclear quadrupolar contribution
determined from recent M\"ossbauer results \cite{Abd}. At 20 mK a maximum
in $C_{el}(T)/T$ marks the onset of
very weak AF order \cite{Custers Nature}. This is suppressed by a tiny
critical magnetic field of $B_c=0.027$ T
applied in the easy plane. At $B=B_c$, a power law divergence
$C_{el}(T)/T \propto T^{-1/3}$ is observed (which is
already incompatible with the 2D-SDW picture) \cite{Custers Nature}.
At higher temperatures the zero-field
specific heat coefficient also varies as $\log(T_0'/T)$ with $T_0'=30$ K
(Fig.~4) \cite{Trovarelli Letter}.
Because of the difference between $T_0'$ and $T_0$, the Gr\"uneisen ratio
is strongly temperature dependent.
Below 1 K it diverges as $\Gamma(T)=\Gamma_0+cT^{-2/3}$, i.e., weaker
than the ${1 \over T}{{\ln\ln(T)} \over
{\ln(T)}}$ form expected in a 2D-SDW picture \cite{Zhu}.
We note that, in the measured temperature range,
the zero-field data of both the specific heat and thermal
expansion are identical to their counterparts at the critical
magnetic field.

To interpret our results, we introduce a Gr\"uneisen exponent $x$ in
terms of the critical
Gr\"uneisen ratio $\Gamma^{cr} \propto {\beta^{cr} \over C_V^{cr}}
\propto {1 \over T^x}$, where $\beta^{cr}$ and
$C_V^{cr}$ are the thermal expansion and specific heat with the
background contributions subtracted;
this exponent
is equal to the dimension of the
most relevant operator that is coupled to
pressure \cite{Zhu}.
It is shown
in Ref.~\cite{Sietal} that, for magnetically three-dimensional
systems without frustration the SDW picture should
apply. This is consistent with our finding here that both the thermal
expansion and specific heat results in
CeNi$_2$Ge$_2$ can be fit by the respective expressions for a 3D-AF-SDW
theory
~\cite{Kadowaki}.
Our results correspond to
$\beta^{cr} \propto \sqrt{T}$ and $C_V^{cr} \propto T^{3/2}$, leading
to $\Gamma_{cr} \propto {1 \over T}$. In
other words, the Gr\"uneisen exponent $x=1$
(with error bars $+0.05/-0.1$,
as determined from a log-log plot shown in the inset
of Fig.~3.)
In an SDW picture,
the most relevant term is
the
quadratic part of the $\phi^4$-theory.
The corresponding dimension
is $1/\nu z =1$, where the
spatial-correlation-length exponent $\nu=1/2$ and the dynamic
exponent $z=2$.

For YbRh$_2$(Si$_{1-x}$Ge$_x$)$_2$, on the other hand, the
measured
Gr\"uneisen exponent is fractional:
$x = 0.7 \pm 0.1$ as determined from a log-log plot of
$\Gamma^{cr}$ versus temperature shown in the inset of Fig.~4.
While
definitely
not compatible with the
itinerant SDW theory, a fractional Gr\"uneisen
exponent is consistent with the locally quantum critical point.
One kind of condition favorable for this new type
of QCP corresponds to a magnetic fluctuation spectrum that is strongly
anisotropic \cite{Sietal}. At such a
locally quantum critical point, spatially local critical excitations
emerge and co-exist with the spatially
extended critical spin fluctuations. There are then two scaling dimensions
to be considered. For the tuning of
the long-wavelength fluctuations, the dimension of interest is still given
by the expression $1/\nu z$. While
$\nu$ remains $1/2$, the dynamic exponent $z$ becomes $2/\alpha>2$
where $\alpha$ is the fractional exponent that
characterizes the dynamical spin susceptibility. As a result, $1/\nu z < 1$.
For the tuning of the local
fluctuations, the corresponding dimension is the inverse of the
temporal-correlation-length exponent. Within an
$\epsilon-$expansion scheme as carried out in Ref.~\cite{Zhu-BFK}
and for the XY-spin-invariant case of relevance
to YbRh$_2$(Si$_{1-x}$Ge$_x$)$_2$, this exponent is found
\cite{Zhu-unpublished}
to be $0.62$ to the first order in
$\epsilon$ and $0.66$ to the second order in $\epsilon$. The overall
Gr\"uneisen ratio will then display a
fractional exponent, as indeed seen experimentally.

We are grateful to
M. Lang, O. Trovarelli and H. Wilhelm
for valuable conversations,
F. Weickert and J. Custers
for their help with the resistivity experiments,
and the Fonds der Chemischen
Industrie (Dresden), the Dutch Foundation FOM-ALMOS (O.T. and J.A.M.),
NSF, TCSAM, and the Welch Foundation
(L.Z. and Q.S.)
for support.

\newpage
\begin{table}
\begin{center}
\begin{tabular}{|l|l|l|} \hline

 $\alpha(T)=a\sqrt{T}+bT$ & $\alpha \parallel c$ & $a=1.5\cdot10^{-6}
$K$^{-1.5}$,$b=0.87\cdot10^{-6}$K$^{-2}$\\
 & $\alpha\parallel a$ & $a=0.99\cdot10^{-6}$K$^{-1.5}$,$b=0.42
\cdot10^{-6}$K$^{-2}$\\
 $\beta(T)=a\sqrt{T}+bT$ & $\beta$ & $a=3.5\cdot10^{-6}$K$^{-1.5}$,
$b=1.7\cdot10^{-6}$K$^{-2}$\\


\hline $C(T)/T=\gamma_0-c\sqrt{T}+d/T^3$ & &$\gamma_0=0.46
$JK$^{-2}$mol$^{-1}$,$c=0.11$Jmol$^{-1}$K$^{-5/2}$\\
 & & $d=102 \mu$JKmol$^{-1}$\\

\hline
\end{tabular}
\end{center}
\caption{\label{a} {\footnotesize
Fit forms
and parameters
for
CeNi$_2$Ge$_2$.
}}
\end{table}

\newpage
\begin{figure}
\centerline{\includegraphics[width=.7\textwidth]{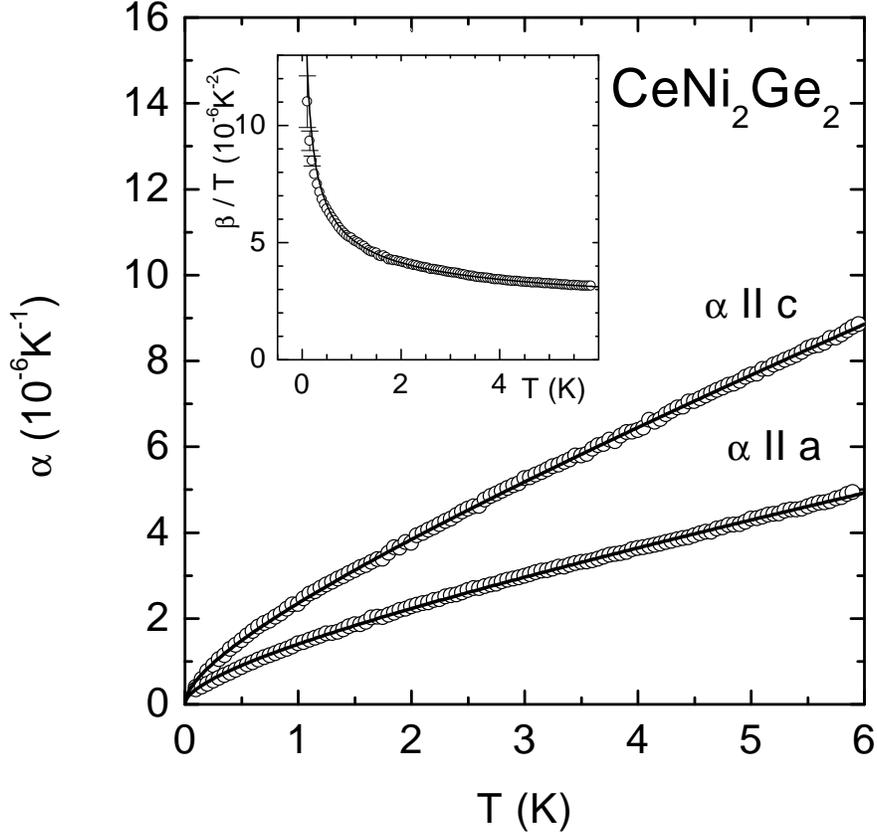}}
\caption{
Linear
thermal expansions of CeNi$_2$Ge$_2$
vs. temperature at
B=0.
Inset shows volume expansion
as ${\beta/T}$ vs $T$.
Solid lines
are fits
as specified in
TABLE I.} \label{fig1}
\end{figure}

\newpage
\begin{figure}
\centerline{\includegraphics[width=.7\textwidth]{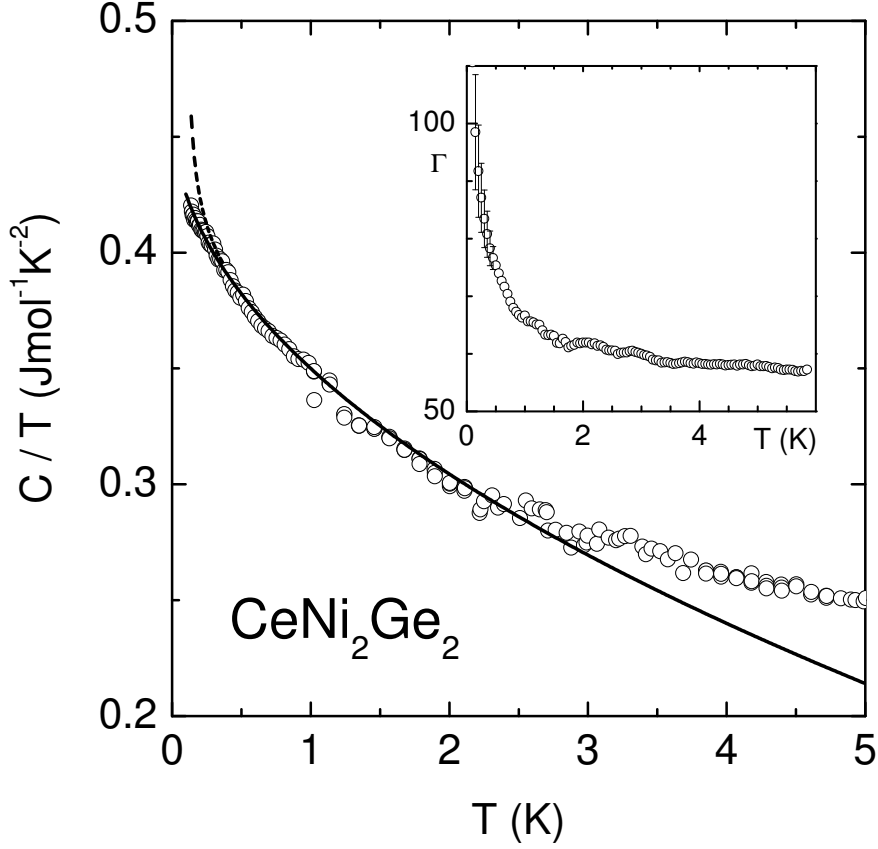}}
\caption{
Specific heat at
B=0
as ${C/T}$ vs $T$
for CeNi$_2$Ge$_2$.
From the raw
data (dashed line at low $T$), a contribution ${C_n=\alpha/T^2}$ with
${\alpha=102}$ $\mu$JK/mol has been
subtracted giving the low-$T$ open circles. The inset shows the
$T$-dependence
of the Gr\"uneisen ratio
${\Gamma=V_{m}/\kappa_T\cdot\beta/C}$ where ${V_m}$
and ${\kappa_T}$ are the molar volume and isothermal
compressibility, respectively. 
Here, we use ${\kappa_T=1.15 \times 10^{-11}}$ Pa$^{-1}$ as 
determined from high-pressure lattice parameter measurements 
at room temperature \cite{Spain}. 
Solid line is a fit
as specified in
TABLE I.} \label{fig2}
\end{figure}

\newpage
\begin{figure}
\centerline{\includegraphics[width=.7\textwidth]{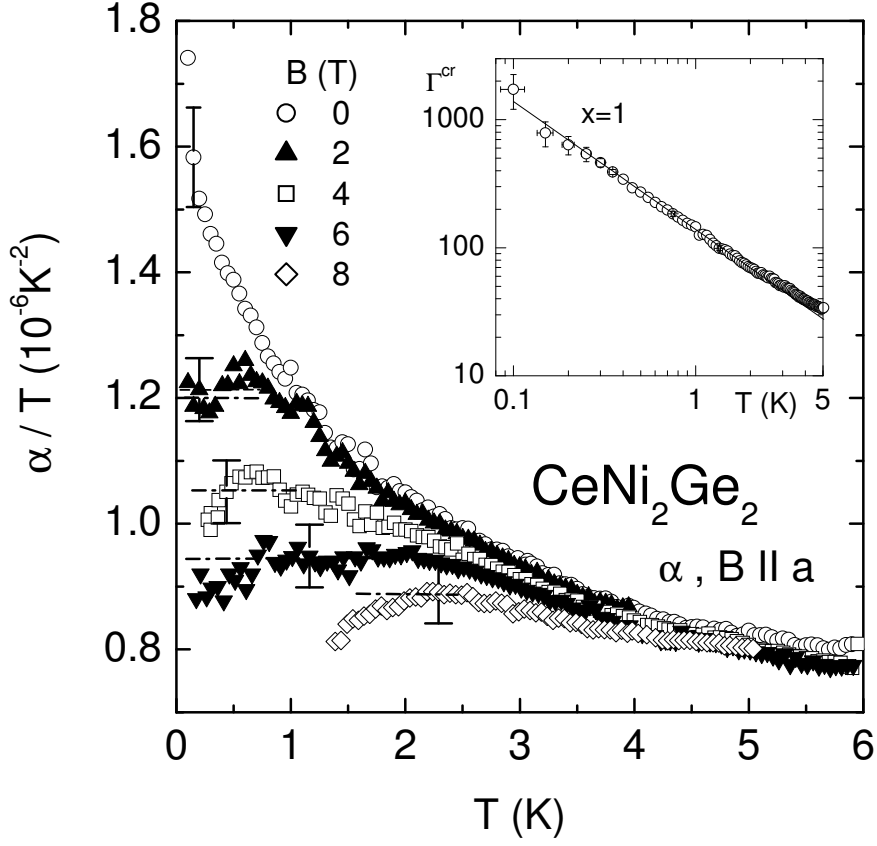}}
\caption{Thermal expansion of CeNi$_2$Ge$_2$
along the $a$-axis as ${\alpha/T}$ vs $T$ at varying magnetic fields.
Inset shows the critical
Gr\"uneisen ratio
$\Gamma^{cr}=V_m/\kappa_T\cdot\beta^{cr}/C^{cr}$ as
$\log\Gamma^{cr}$ {\it vs} $\log T$ (at $B=0$) with
$\beta^{cr}=\beta(T)-bT$ and $C^{cr}=C(T)-(\gamma T+d/T^2)$
using the parameters listed in TABLE I. Solid line represents
$\Gamma^{cr}\propto 1/T^x$ with $x=1$.} \label{fig3}
\end{figure}

\newpage
\begin{figure}
\centerline{\includegraphics[width=.7\textwidth]{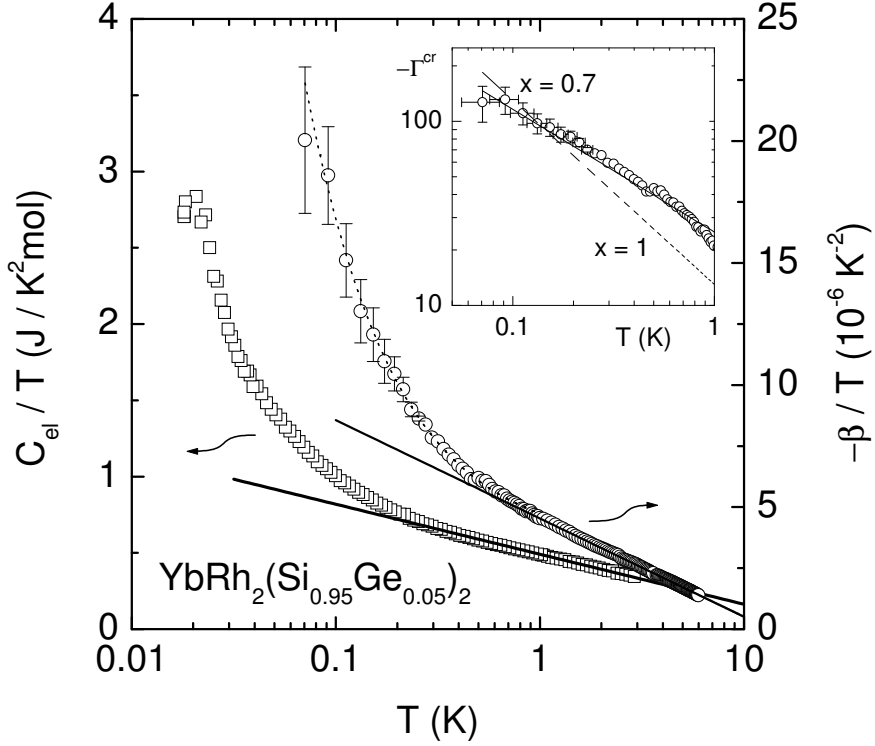}}
\caption{Electronic specific heat as
${C_{el}/T}$
(left axis) and volume thermal expansion
as ${-\beta/T}$ (right axis) vs $T$ (on a
logarithmic scale) for YbRh$_2$(Si$_{0.95}$Ge$_{0.05}$)$_2$ at
B=0.
Solid lines indicate
${\log(T_0/T)}$ dependences with ${T_0=30}$ K and 13 K for ${C_{el}/T}$
and ${-\beta/T}$, respectively. Dotted
line represents $-\beta/T=a_0+a_1/T$ with $a_0=3.4\cdot10^{-6}$ K$^{-2}$
and $a_1=1.34\cdot10^{-6}$ K$^{-1}$.
Inset displays the log-log plot of $\Gamma^{cr}(T)$ with
$\Gamma^{cr}=V_m/\kappa_T\cdot\beta^{cr}/C^{cr}$ using
${\kappa_T=5.3\cdot10^{-12}}$ Pa$^{-1}$ \cite{Abd},
$\beta^{cr}=\beta(T)+a_0T$ and $C^{cr}=C_{el}(T)$. Solid and
dotted lines represent $\Gamma^{cr}\propto 1/T^x$ with $x=0.7$
and $x=1$, respectively.} \label{fig4}
\end{figure}


\begin{references}
\bibitem{SC CNG} A superconducting transition, observed below 0.2 K in
some electrical resistivity measurements \cite{Grosche,Gegenwart
Physica,Koerner,Braithwaite}, has not been detected by any bulk
probe so far.
\bibitem{YRS beta}The data for $T<1$ K allow a fitting with
$a_1 + a_0T + a_2\ln \ln(T_0/T)$ \cite{Zhu}, but the extracted
$\ln\ln(T_0/T)$ component is negligible compared to the other
terms.
\end{references}
\end{document}